\documentclass[a4paper,12pt]{article}
\usepackage{amssymb}
\usepackage{amsthm}
\usepackage{amsmath}

\usepackage{latexsym}
\makeatletter

\newdimen\LENB \newdimen\LENW \newdimen\THI
\newdimen\LENWH \newdimen\LENTOT \newcount\N
\def\vbrknlnele#1#2#3{
  \LENB=#1pt \LENW=#2pt \THI=#3pt
  \LENWH=\LENW \divide\LENWH by 2
  \LENTOT=\LENB \advance\LENTOT by \LENW
  \vbox to \LENTOT{
    \vbox to \LENWH{}
    \nointerlineskip
    \vbox to \LENB{\hbox to \THI{\vrule width \THI height \LENB}}
    \nointerlineskip
    \vbox to \LENWH{}
  }}

\def\vbrknln#1{
  \N=#1
  \vcenter{
    \vbox{
      \loop\ifnum\N>0
        \vbox to 4pt{\vbrknlnele{2}{2}{0.1}}
        \nointerlineskip
        \advance\N by -1
      \repeat
  }}}

\def\vbl#1{\hskip-5pt \vbrknln{#1} \hskip-5pt}

\def\hbrknlnele#1#2#3{
  \LENB=#1pt \LENW=#2pt \THI=#3pt
  \LENTOT=\LENB \advance\LENTOT by \LENW
  \vcenter{
    \vbox to \THI{
      \hbox to \LENTOT{
        \hfil
        \vrule width \LENB height \THI
        \hfil}
  }}}

\def\hblele{\hbrknlnele{2}{2.2}{0.1}}

\def\hblfil{\cleaders\hbox{$ \m@th \mkern1mu \hblele \mkern1mu
$}\hfill} \makeatother

\author{
Wen-Xiu Ma\thanks{Email: mawx@math.usf.edu}
\\{ \small Department of Mathematics, University of South Florida,
Tampa, FL 33620-5700, USA} \vspace{2mm} \\
Ken-ichi Maruno\thanks{Email: maruno@math.kyushu-u.ac.jp}
\\
{\small Faculty of Mathematics, Kyushu University,}\\
{\small  Hakozaki, Higashi-ku, Fukuoka, 812-8581, Japan
} }
\title
{\sf Complexiton Solutions of the Toda Lattice Equation}
\date{\nonumber}
\numberwithin{equation}{section}

\begin{document}
\maketitle

\newcommand{\R}{\mathbb{R}}
\newcommand{\C}{\mathbb{C}}
\newcommand{\D}{\displaystyle}

\begin{abstract}

\setlength{\baselineskip}{18pt}

A set of coupled conditions consisting of differential-difference
equations is presented for Casorati determinants to solve the Toda
lattice equation. One class of the resulting conditions leads to
an approach for constructing complexiton solutions to the Toda
lattice equation through the Casoratian formulation. An analysis
is made for solving the resulting system of
differential-difference equations, thereby providing the general
solution yielding eigenfunctions required for forming
complexitons. Moreover, a feasible way is presented to compute the
required eigenfunctions, along with examples of real complexitons
of lower order.

\vskip 6mm




\vskip 1cm

\end{abstract}

\newtheorem{thm}{Theorem}[section]
\newtheorem{lem}{Lemma}[section]
\newtheorem{prop}{Proposition}[section]
\newtheorem{defi}{Definition}[section]


\setlength{\baselineskip}{17.6pt}
\def \part {\partial}
\def \be {\begin{equation}}
\def \ee {\end{equation}}
\def \bea {\begin{eqnarray}}
\def \eea {\end{eqnarray}}
\def \ba {\begin{array}}
\def \ea {\end{array}}
\def \si {\sigma}
\def \al {\alpha}
\def \la {\lambda}

\newpage
\section{Introduction}

Many integrable equations, both continuous and discrete, have
soliton solutions, which are exponentially decaying at spatial
infinity. The existence of a three-soliton solution often
indicates the integrability of the equation under investigation.
The Toda lattice equation is one of the well-known lattice model
equations exhibiting the soliton phenomenon \cite{Toda-JPSJ1967}.
Its multi-soliton solutions can be expressed through Casorati
determinants \cite{Hirota-JPSJ1973,Nimmo-PLA1983,Hirota-2DTL1988},
and its approximate soliton solutions have been explored around an
exact soliton solution \cite{TakenoKH-PLA1988}.

There are also positons and negatons to the Toda lattice equation,
which can be presented by generalized Casorati determinants
\cite{StahlhofenM-JPA1995,MarunoMO-JPSJ2003}, and solitons are
just a specific class of negatons. Positons and negatons were
first presented for the Korteweg-de Vries (KdV) equation (see, for
example, \cite{Matveev-PLA1992,RasinariuSK-JPA1996}). All three
classes of solutions--solitons, positons and negatons--are
associated with real eigenvalues of the associated spectral
problems. Moreover, the absolute values of soliton and negaton
solutions contain one kind of elementary transcendental functions
- exponential functions of the space variables, and the absolute
values of positon solutions contain another kind of elementary
transcendental functions - trigonometrical functions of the space
variables.

A challenging problem in solution theory is how to construct a
different kind of explicit exact solutions to soliton equations,
whose absolute values involve both exponential and trigonometrical
functions of the space variables and which are associated with the
complex eigenvalues of the associated spectral problems. The
absolute values do not need to be taken if solutions are real, as
in the case of the KdV equation; but do need to be taken if
solutions are complex, as in the case of the nonlinear
Schr\"odinger equation. Exact solutions of such kind are called
complexiton solutions and have been presented for the KdV equation
\cite{Ma-PLA2002}. Note that interaction solutions between
positons and negatons can also contain both exponential and
trigonometrical functions of the space variables, but they are
associated with real eigenvalues of the associated spectral
problems and thus they are not examples of so-called complexitons.
Although these solutions belong to a broader class of exact
solutions, they can be well formulated once three kinds of basic
solutions--negatons, positons and complexitons--are presented
\cite{MaY-TAMS2003}.

Therefore, for the Toda lattice equation, the basic question for
us is whether there exist complexiton solutions and how one can
construct complexitons if they exist. This is the topic that we
would like to address in this paper. It is known that the
Casoratian formulation is a powerful technique to generate
explicit solutions of integrable lattice equations
\cite{Nimmo-PLA1983,Hirota-2DTL1988}. Solutions determined by the
Casorati determinant technique and generalized Casorati
determinant technique are called Casorati determinant solutions
and generalized Casorati determinant solutions, respectively
\cite{MarunoMO-JPSJ2003}. For the Toda lattice equation, solitons
are examples of Casorati determinant solutions
\cite{Hirota-JPSJ1973,Nimmo-PLA1983}, and positons and nagatons
are examples of generalized Casorati determinant solutions
\cite{MarunoMO-JPSJ2003}.

In this paper, we would like to show that there exist complexiton
solutions of the Toda lattice equation through the Casoratian
formulation. Inspired by its Lax pair, a set of coupled conditions
will be presented for guaranteeing Casorati determinants to be
solutions of the Toda lattice equation in Section 2, and this
yields an approach to a broad class of Casorati determinant
solutions and generalized Casorati determinant solutions of the
Toda lattice equation. The resulting coupled conditions will be
used to construct real complexiton solutions to the Toda lattice
equation in Section 3. Moreover, a feasible way will be proposed
to construct sets of special eigenfunctions satisfying the
required conditions in Section 4, together with concrete examples
of real complexitons of lower order. A few concluding remarks will
be given in Section 5.

\section{Casoratian formulation}
 \label{sec:casorati:toda-c}

Let us consider the Toda lattice equation in the following form:
\begin{equation}
\dot{a}_n=a_n(b_{n-1}-b_n)\,,\quad
\dot{b}_n=a_n-a_{n+1}\,,\label{eq:toda:toda-c}
\end{equation}
where (also in the rest of the paper) the dot denotes the
differentiation with respect to the time variable $t$. This Toda
lattice equation can be reduced to the periodic case
($a_{n+N}=a_n$ and $ b_{n+N}=b_n$ for some positive integer $N$)
and the finite case (only finitely many $a_n$ and $b_n$ are
non-zero). It is also more general than the square form of the
Toda lattice equation \cite{Toda-book1989}:
\begin{equation}
\dot{a}_n=a_n(b_{n}-b_{n+1})\,,\quad
\dot{b}_n=2(a_{n-1}^2-a_{n}^2)\,,\label{eq:squareformoftoda:toda-c}
\end{equation}
because there is a solution transformation
\[(a_n(t),b_n(t))\to \bigl((a_{n-1}(\frac 12 t))^2,b_n(\frac 12 t)\bigr)\]
from the square form (\ref{eq:squareformoftoda:toda-c}) to the
non-square form (\ref{eq:toda:toda-c}). On the other hand, the
Toda lattice equation (\ref{eq:toda:toda-c}) is the isospectral
$(\lambda _t=0$) compatibility condition of the following spectral
problems: \be \left\{\ba {l} \dot \phi (n)=b_{n-1}\phi(n)
+\phi(n-1),
\vspace{2mm}\\
a_n \phi(n+1)+b_{n-1}\phi(n) +\phi(n-1)=\lambda \phi(n), \ea
\right.\label{eq:spoftoda:toda-c}
 \ee
where $\lambda $ is a spectral parameter. Namely, it has the Lax
representation: \be \dot L=[A,L], \ee where the Lax pair is
defined by
\begin{equation} \left\{ \ba {l}
L_{nm}
=a_n\delta_{n+1,m}+b_{n-1}\delta_{nm}+\delta_{n-1,m},\vspace{2mm} \\
A_{nm}=\delta _{n+1,m}+b_{n-1}\delta _{nm}. \ea \right.
\end{equation}

 Under an dependent variable
transformation
\begin{equation}
a_n=1+\frac{d^2}{dt^2}\log
\tau_n=\frac{\tau_{n+1}\tau_{n-1}}{\tau_n^2}\,, \quad
b_n=\frac{d}{dt}\log \frac{\tau_{n}}{\tau_{n+1}}=\frac {\dot
{\tau}_{n}\tau_{n+1}-\tau_n\dot{\tau}_{n+1}
}{\tau_n{\tau}_{n+1}}\,, \label{eq:todavariable:toda-c}
\end{equation}
we have
\[ \left\{ \ba {l}
\dot{a}_n-a_n(b_{n-1}-b_n)=0,\vspace{2mm}\\
\D \dot{b}_n-a_n+a_{n+1}= \frac {\ddot{\tau}_n\tau
_n-(\dot{\tau}_n)^2-\tau_{n+1}\tau_{n-1}+\tau_n^2}{\tau _n^2}
-\frac {\ddot{\tau}_{n+1}\tau _{n+1}
-(\dot{\tau}_{n+1})^2-\tau_{n+2}\tau_{n}+\tau_{n+1}^2}{\tau_{n+1}^2},
\ea \right .\] and thus
 the Toda lattice equation (\ref{eq:toda:toda-c}) can be
 satisfied if we require
the bilinear equation
\begin{equation}
 [\frac 12 D_t^2 -2 \sinh ^2(\frac {D_n}2 )]\tau_n \cdot \tau_n=
 \ddot{\tau}_n\tau _n-(\dot{\tau}_n)^2-\tau_{n+1}\tau_{n-1}+\tau_n^2=0\,,
\label{eq:bltl:toda-c}
\end{equation}
where $D_t$ and $D_n$ are Hirota's operators. This is called the
bilinear Toda lattice equation. Through the dependent variable
transformation (\ref{eq:todavariable:toda-c}), multi-soliton
solutions of the Toda lattice equation (\ref{eq:toda:toda-c}) can
be presented by the Casorati determinant
\cite{Nimmo-PLA1983,MatveevS-book1991}: \be {\rm
Cas}(\phi_1(n),\phi_2(n),\cdots,\phi_{N}(n)) :=\left|
\ba {cccc}\phi _1(n)&\phi _1(n+1)&\cdots &\phi_1(n+N-1)\\
\phi _2(n)&\phi_2(n+1)&\cdots &\phi_2(n+N-1)
\\ \vdots &\vdots & \ddots & \vdots \\
\phi _{N}(n)&\phi_{N}(n+1)&\cdots &\phi_N(n+N-1) \ea \right |,\
N\ge 1\,, \label{Casoratian} \ee provided that the functions
$\phi_i(n)$, $1\le i\le N$, solve
\begin{equation}
\phi_i(n+1)+\phi_i(n-1)=\lambda _i \phi_i(n)\, ,\quad
(\phi_{i}(n))_t=\phi_i(n-1)\, , \quad 1\le i\le N\,
 \label{eq:solitoncondition:toda-c}
\end{equation}
where $\lambda _i=2\,\textrm{cosh}(k_i) $ and the $k _i$'s are
arbitrary distinct real constants. The conditions in
(\ref{eq:solitoncondition:toda-c}) correspond to the case of the
spectral problems (\ref{eq:spoftoda:toda-c}) with $a_n=1$ and
$b_n=0$, a special solution to the Toda lattice equation
(\ref{eq:toda:toda-c}). This also implies that the Casorati
determinant solution is actually resulted from the Darboux
transformation of the Toda lattice equation.

In what follows, we would like to show that the Casorati
determinant presents a very broad class of exact solutions to the
Toda lattice equation (\ref{eq:toda:toda-c}), among which
solitons, positons and negatons are special examples. The
following theorem is a generalization to the cases of solitons,
positons and negatons.

\begin{thm} Assume that
a set of functions $\phi_i(n)=\phi_i(n,t), 1\le i\le N$, solve a
system of differential-difference equations
\begin{eqnarray}
&&\phi_i(n+1)+\phi_i(n-1)=\sum_{j=1}^N \lambda _{ij}
\phi_j(n)\,,\quad 1\le i\le N\,,
\label{eq:detrelationforn:toda-c}\\
&&\partial_t \phi_{i}(n)=\phi_i(n+\delta )\,,\quad 1\le i\le
N\,,\label{eq:detrelationfort:toda-c}
\end{eqnarray}
where $\delta =\pm 1 $ and the $\lambda_{ij}$'s are arbitrary
constants.
 Then the Casorati determinant
  \[\tau_n={\rm
Cas}(\phi_1(n),\cdots,\phi_N(n))\] gives a solution to the
bilinear Toda lattice equation (\ref{eq:bltl:toda-c}), and further
the dependent variable transformation
(\ref{eq:todavariable:toda-c}) presents a solution to the Tada
lattice equation (\ref{eq:toda:toda-c}).
\label{thm:generalconditionsforCas:toda-c}
\end{thm}

\begin{proof}
We only prove the result under (\ref{eq:detrelationfort:toda-c})
with $\delta =-1$. The other case with $\delta =1$ is completely
similar. Assuming that \be \Phi_N(n)=(\phi_1(n),\cdots
,\phi_N(n))^T,\label{eq:defofPhi_N:toda-c} \ee
 we adopt the notation
\begin{equation}
|i_1,i_2,\cdots ,
i_N|:=\textrm{det}(\Phi_N(n+i_1),\Phi_N(n+i_2),\cdots
,\Phi_N(n+i_N))\,,\  {k..l}:={k,k+1,\cdots,l}\,,
\end{equation}
where $i_1,i_2,\cdots,i_N$ and $k< l$ are arbitrary integers. For
example, we have
 \[\ba{l}
  |0..N-2,N|=
 \textrm{det}(\Phi_N(n),\Phi_N(n+1),\cdots,\Phi_N(n+N-2),\Phi_N(n+N)),\vspace{2mm}\\
|-1,1..N-1|=\textrm{det}(\Phi_N(n-1),\Phi_N(n+1),\Phi_N(n+2),\cdots,\Phi_N(n+N-1)).\ea
 \]

 Directly from the conditions in (\ref{eq:detrelationfort:toda-c}) with $\delta =-1$, we obtain
the expressions for the first two derivatives of the
$\tau$-function $\tau_n$ with respect to $t$:
  \begin{equation}
\dot{\tau}_n=|-1,1..{N-1}| \,,\
\ddot{\tau}_n=|-2,1..N-1|+|-1,0,2..N-1| \,.
\label{eq:taurelation-tandtt:toda-c}\end{equation} On the other
hand, we have the general result for any determinant $|A_{ij}|$:
\be  \sum_{k=1}^N|A_{ij}|_k=\sum_{k=1}^N|A_{ij}|^k\,,
\label{eq:equalitiesfordetofA_{ij}:toda-c} \ee where $|A_{ij}|_k$
denotes the determinant $A_{ij}$ with its $k$-th row operated by
the operator $S$: \be (S\phi)(n):=\phi(n+1)+\phi(n-1),
\label{eq:defforS:toda-c} \ee
 and $|A_{ij}|^k$ denotes the
determinant with its $k$-th column operated by the operator $S$.
Applying (\ref{eq:equalitiesfordetofA_{ij}:toda-c}) to two
determinants $|0..N-1|$ and $|-1,1..N-1|$ and using the conditions
in (\ref{eq:detrelationforn:toda-c}), we obtain the determinant
identities: \begin{eqnarray} &&\sum_{i=1}^N\lambda_{ii}|0..N-1|
=|0..N-2,N|+|-1,1..N-1|
\,,\label{eq:taurelation1:toda-c}\\
&&\sum_{i=1}^N\lambda_{ii}|-1,1..N-1|=|0..N-1|+|-2,1..N-1|
\nonumber\\
&&+|-1,0,2..N-1| +|-1,1..N-2,N| \,.\label{eq:taurelation2:toda-c}
\end{eqnarray}

Now, making use of (\ref{eq:taurelation-tandtt:toda-c}),
(\ref{eq:taurelation1:toda-c}) and (\ref{eq:taurelation2:toda-c}),
we find that the left-hand side of (\ref{eq:bltl:toda-c}) gives
the terms \begin{eqnarray} &&\ddot{\tau}_n\tau
_n-(\dot{\tau}_n)^2-\tau_{n+1}\tau_{n-1}+\tau_n^2
\nonumber\\
&&=\bigl(|-2,1..N-1|+|-1,0,2..N-1|\bigr)|0..N-1|-|-1,1..N-1|^2\nonumber\\
&& \quad -|1..N||-1..N-2| +|0..N-1|^2\nonumber\\
&& =\bigl(\sum_{i=1}^N\lambda_{ii} |-1,1..N-1|-|0..N-1| -|-1,1..N-2,N|\bigr)|0..N-1|\nonumber\\
&& \quad -\bigl(\sum_{i=1}^N\lambda_{ii}|0..N-1|-|0..N-2,N|\bigr)|-1,1..N-1| \nonumber\\
   & & \quad -|1..N||-1..N-2| + |0..N-1| ^2\nonumber\\
&& =-|-1,1..N-2,N||0..N-1|+|0..N-2,N||-1,1..N-1|-|1..N||-1..N-2|\,
.\qquad\nonumber
\end{eqnarray}
The last sum above is the Laplace expansion by ${\it N} \times
{\it N}$ minors of the following $2{\it N}\times 2{\it N}$
determinant
\[ -\frac 12 \left \vert
\ba {cccccccc} 1..N-2 & \vbl{4}& \hbox{\O} &\vbl{4} & -1&0 &N-1& N \vspace{0.8mm}\\
\multispan{8} \hblfil \\
 \hbox{\O}& \vbl{4} & 1..N-2 & \vbl{4} & -1&0 &N-1& N  \ea
\ \right \vert , \] where $\hbox{\O}$ indicates the $N\times(N-2)$
zero matrix. This can be easily shown to be identically zero.
Thus, the solution is verified.
\end{proof}

Note that the first half conditions in
(\ref{eq:solitoncondition:toda-c}) are just a special case of the
conditions in (\ref{eq:detrelationforn:toda-c}). Therefore, we can
expect to get more solutions to the Toda lattice equation
(\ref{eq:toda:toda-c}) by solving the system of
differential-difference equations,
(\ref{eq:detrelationforn:toda-c}) and
(\ref{eq:detrelationfort:toda-c}), as in the KdV case
\cite{MaY-TAMS2003}. Moreover, the entire problem of constructing
explicit solutions is reduced to the problem of solving the
system, (\ref{eq:detrelationforn:toda-c}) and
(\ref{eq:detrelationfort:toda-c}).

The system of (\ref{eq:detrelationforn:toda-c}) and
(\ref{eq:detrelationfort:toda-c}) can be compactly written as \be
(S\Phi_N)(n)\equiv \Phi_N(n+1)+\Phi_N(n-1)=\Lambda \Phi_N (n),\
(\Phi_N(n))_t=\Phi_N(n+\delta ),
\label{eq:compactformoflinearconsditionsforphi_i:toda-c} \ee where
$\Phi_N$ is defined by (\ref{eq:defofPhi_N:toda-c}) and \be
\Lambda: =(\lambda _{ij})_{N\times N}\ee is called the coefficient
matrix of the system of (\ref{eq:detrelationforn:toda-c}) [or the
system of (\ref{eq:detrelationforn:toda-c}) and
(\ref{eq:detrelationfort:toda-c})]. Note that a constant similar
transformation for the coefficient matrix $\Lambda $ does not
change the resulting Casorati determinant solution to the Toda
lattice equation (\ref{eq:toda:toda-c}). Actually, if we have $M
=P^{-1}\Lambda P$ for some constant invertible matrix $P$, then
$\tilde \Phi_N=P\Phi_N$ satisfies \be (S\tilde \Phi_N)(n)\equiv
\tilde \Phi_N(n+1)+\tilde \Phi_N(n-1)=M \tilde \Phi_N (n),\
(\tilde \Phi_N(n))_t=\tilde \Phi_N(n+\delta ). \ee Obviously, the
dependent variable transformation (\ref{eq:todavariable:toda-c})
leads to the same Casorati determinant solutions from $\Phi_N$ and
$\tilde \Phi_N$. Therefore, by linear algebra, we only need to
consider the following two types of Jordan blocks of the
coefficient matrix $\Lambda$: \bea && \left[\ba {cccc}\lambda_j &
& & 0
\vspace{2mm}\\
  1 & \lambda_j  &  &  \vspace{2mm}\\
\vdots &\ddots& \ddots&  \vspace{2mm}\\
 0 &\cdots  & 1 & \lambda_j  \ea \right]_{k_j\times k_j},
\label{eq:1stJordanblock:toda-c}
\\
&& \left[\ba {cccc} A_j & & & 0
\vspace{2mm}\\
   I_2 & A_j& & \vspace{2mm}\\
\vdots & \ddots& \ddots& \vspace{2mm}\\
  0 &\cdots &I_2 &A_j \ea \right]_{l_j\times l_j},\ A_j=\left[\ba
{cc}\alpha _j&-\beta _j\vspace{2mm}\\ \beta _j&\alpha _j \ea
\right],\   I_2=\left[\ba {cc}1 &0\vspace{2mm}\\
0&1 \ea \right],
 \label{eq:2rdJordanblock:toda-c} \eea where
$\lambda_j$,  $\alpha_j $ and $\beta_j >0$ are all real constants,
and $k_j$ and $l_j$ are positive integers. The first type of
Jordan blocks has the real eigenvalue $\lambda_j$ with algebraic
multiplicity $k_j$, and the second type of Jordan blocks has the
complex eigenvalues $\lambda _{j,{\pm}} =\alpha_j\pm \beta _j i$
with algebraic multiplicity $l_j$.

The case of real eigenvalues corresponds to positons, nagatons and
rational solutions \cite{MarunoMO-JPSJ2003,MaY-CSF2004}. In what
follows, we will focus on the case of complex eigenvalues to
present complexitons. We will show how to solve the system of
differential-difference equations,
(\ref{eq:detrelationforn:toda-c}) and
(\ref{eq:detrelationfort:toda-c}), in the case of complex
eigenvalues such that the Casoratian formulation leads to real
complexiton solutions of the Toda lattice equation.

\section{Complexiton solutions}
\label{sec:structureofcomplexitons:toda-c}

In order to construct complexitons, let us begin to solve the
system of differential-difference equations,
(\ref{eq:detrelationforn:toda-c}) and
(\ref{eq:detrelationfort:toda-c}), whose coefficient matrix
consists of Jordan blocks of the second type. Since all subsystems
corresponding to Jordan blocks are separated, it suffices to
illustrate how to solve a system associated with one Jordan block
of the second type. Let us specify such a system as \be (
S\Phi)(n)\equiv \Phi(n+1)+\Phi(n-1) =\Lambda \Phi(n) , \
(\Phi(n))_t=\Phi(n+\delta )
,\label{eq:representativesysytemforcomplexitonsol:toda-c}\ee where
$\delta =\pm 1$
and \be \Phi= \left[\ba {c} \phi_1\vspace{2mm}\\
\phi_2\vspace{2mm}\\ \vdots \vspace{2mm}\\ \phi_{2l-1}\vspace{2mm}\\
\phi_{2l}\ea \right] ,\ \Lambda = \left[\ba {cccc} A & & & 0
\vspace{2mm}\\
   I_2 & A& & \vspace{2mm}\\
\vdots & \ddots& \ddots& \vspace{2mm}\\
  0 &\cdots &I_2 &A \ea \right]_{l\times l},\ A=\left[\ba
{cc}\alpha &-\beta \vspace{2mm}\\ \beta &\alpha  \ea \right]. \ee
If we introduce \be \lambda =\alpha +\beta i,\
F_k=\phi_{2k-1}+\phi_{2k}i,\  1\le k\le l, \ee then the system
(\ref{eq:representativesysytemforcomplexitonsol:toda-c}) is
obviously equivalent to the following triangular system for all
$F_k$: \be SF_{k} =\lambda F_{k}+F_{k-1}, \ (F_{k}(n))_t=F_{k}
(n+\delta ) ,\ 1\le k\le l,
\label{eq:compactrepresentativesysytemforcomplexitonsol:toda-c}
 \ee where $F_0=0$.

\begin{lem}\label{le:gsof2ndtypesystem:toda-c} Let $\lambda $ be a complex number not equal 2 and $\delta =\pm 1$.
 Then the
homogeneous system of differential-difference equations \be
(Sf)(n)\equiv f (n+1)+f (n-1)=\lambda f (n), \ (f (n))_t=f
(n+\delta ) , \label{eq:gnlinearconditions:toda-c} \ee has its
general solution \be f(\lambda;c,d)(n)\equiv f(n) = c \omega ^{ n}
\textrm{e}^{ t\omega ^\delta}+ d \omega ^{- n} \textrm{e}^{t\omega
^{-\delta }}, \label{eq:generalsolforgnlinearconditions:toda-c}
\ee where $\omega $ is defined by \be \lambda =\omega +\omega
^{-1}, \ \textrm{i.e.}, \ \omega ^2-\lambda \omega
+1=0,\label{eq:defofomega:toda-c} \ee and $c$ and $d$ are
arbitrary constants.
\end{lem}

\begin{proof}
Note that the general solution to the linear difference equation
\be (Sf)(n)\equiv f(n+1)+f(n-1)=\lambda f(n)=(\omega +\omega
^{-1})f(n) \label{eq:differenceequationforphi:toda-c} \ee has two
free parameters. Moreover, it is easy to show that $f(n)=\omega ^{
n} $ and $f(n)= \omega ^{- n}$ are two solutions to
(\ref{eq:differenceequationforphi:toda-c}), and they are linearly
independent since $\lambda \ne 2$. Hence, the general solution to
the difference equation (\ref{eq:differenceequationforphi:toda-c})
is given by
\[f(n) = c_1(t) \omega ^{ n} + d_1(t ) \omega ^{- n}, \]
where $c_1$ and $d_1$ are two functions of $t$. On the other hand,
the differential equation $(f(n))_t= f(n+\delta )$ requires that
\[ \dot c_1=c_1 \omega ^\delta  ,\
\dot d_1=d_1 \omega ^{-\delta } ,\] and thus we have
\[c_1= c \textrm{e}^{ t\omega ^\delta},\ d_1= d
\textrm{e}^{t\omega ^{-\delta }},  \] where $c$ and $d$ are
arbitrary constants. Therefore, the general solution to the
differential-difference equation
(\ref{eq:gnlinearconditions:toda-c}) is given by
(\ref{eq:generalsolforgnlinearconditions:toda-c}). The proof is
finished.
\end{proof}

\noindent {\bf Remark:} The condition $\lambda = 2$ is equivalent
to $\omega =1$, and moreover, $\omega ^n$ and $\omega ^{-n}$ are
linearly dependent if and only if $\omega =1$. Therefore, $\lambda
\ne 2$ is necessary for guaranteeing the linear independence of
$\omega ^n$ and $\omega ^{-n}$. Actually, the case of $\lambda =2$
corresponds to rational solutions \cite{MaY-CSF2004}.

\begin{thm}\label{thm:solutionformulaforeigenfunctions:toda-c}
 Let $\lambda $ be a complex number not equal 2 and $\delta =\pm 1$.
 Suppose that $f(\lambda ;c,d)$ is the general solution to (\ref{eq:gnlinearconditions:toda-c}), defined by
(\ref{eq:generalsolforgnlinearconditions:toda-c}), and for each
$1\le k\le l$, define $f_k=f(\lambda ;c_k,d_k)$ with a pair of
arbitrary constants $c_k$ and $d_k$. Then the general solution to
the triangular system of differential-difference equations
(\ref{eq:compactrepresentativesysytemforcomplexitonsol:toda-c}) is
given by \be F_k=\sum_{p=0}^{k-1}\frac 1 {p!} \frac {\partial ^p
f_{k-p}}{\partial \lambda ^{p}}= \sum_{p=0}^{k-1}\frac 1 {p!}
\frac {\partial ^p f(\lambda;c_{k-p},d_{k-p})}{\partial \lambda
^{p}} , \ 1\le k\le l.
\label{eq:generalsolutionofcompactrepresentativesysytem:toda-c}
\ee
\end{thm}

\begin{proof}
First, from (\ref{eq:gnlinearconditions:toda-c}), we have
\[ Sf_k=\lambda f_k, \  (f_k(n))_t=f_k(n+\delta ),\ 1\le k\le l. \]
Differentiating these equalities $p$ times with respect to
$\lambda $, we obtain
  \be \left\{\ba {l}
\D S(\frac 1 {p!} \frac {\partial ^p f_{k}}{\partial \lambda
^{p}}) =\lambda (\frac 1 {p!} \frac {\partial ^p f_{k}}{\partial
\lambda ^{p}})+ \frac 1 {(p-1)!} \frac {\partial ^{p-1}
f_{k}}{\partial
\lambda ^{p-1}},\ 1\le k\le l,\ p\ge 1,\vspace{2mm}\\
\D  \bigl[(\frac 1 {p!} \frac {\partial ^p f_{k}}{\partial \lambda
^{p}})(n)\bigr]_t=(\frac 1 {p!} \frac {\partial ^p f_{k}}{\partial
\lambda ^{p}})(n+\delta ) ,\ 1\le k\le l,\  p\ge 1.\ea \right.
\label{eq:ktimesderivativeofF_k:toda-c}\ee
 Second, note
that the linear system
(\ref{eq:compactrepresentativesysytemforcomplexitonsol:toda-c}) is
triangular, and so we can solve the system one by one from $F_1$
through $F_l$.

By Lemma \ref{le:gsof2ndtypesystem:toda-c}, the general solution
to the first subsystem for $F_1$ in
(\ref{eq:compactrepresentativesysytemforcomplexitonsol:toda-c})
can be defined by $f_1$ with a pair of arbitrary constants $c_1$
and $d_1$. Now let $2\le k\le l$ (if $l=1$, we are finished).
Assume that we already solve the first $k-1$ subsystems for $F_p$,
$1\le p\le k-1$. Then the $k$-th subsystem for $F_k$ in
(\ref{eq:compactrepresentativesysytemforcomplexitonsol:toda-c})
can be viewed as a non-homogeneous linear system and thus its
general solution is determined by
\[F_k=F_{k,h}+F_{k,s},\]
where $F_{k,h}$ is the general solution to the homogeneous
counterpart of the $k$-th subsystem and $F_{k,s}$ is a special
solution to the non-homogeneous $k$-th subsystem. Again by Lemma
\ref{le:gsof2ndtypesystem:toda-c}, the general solution $F_{k,h}$
to the $k$-th subsystem of
(\ref{eq:compactrepresentativesysytemforcomplexitonsol:toda-c})
can be defined by $f_k$ with a pair of arbitrary constants $c_k$
and $d_k$.
 On the other hand, by an inspection, a
special solution $F_{k,s}$ to the $k$-th subsystem can be found to
be
\[ F_{k,s} = \D \sum_{p=1}^{k-1}\frac 1 {p!}\frac {\partial ^p
f_{k-p}}{\partial \lambda ^p}. \] This can be proved by using
(\ref{eq:ktimesderivativeofF_k:toda-c}). Actually, we have
\[ \ba {l}
\D (S-\lambda )F_{k,s}=\sum_{p=1}^{k-1} (S-\lambda ) \bigl(\frac 1
{p!}\frac {\partial ^p f_{k-p}}{\partial \lambda ^p}\bigr)
\vspace{2mm}\\
\D  = \sum_{p=1}^{k-1} \frac 1 {(p-1)!}\frac {\partial ^{p
-1}f_{k-p}}{\partial \lambda ^{p-1}} =\sum_{p=0}^{k-2} \frac 1
{p!}\frac {\partial ^p f_{k-p-1}}{\partial \lambda ^p}
=F_{k-1},\vspace{2mm}\\
\bigl(F_{k,s}(n)\bigr)_t = \D \sum_{p=1}^{k-1}\bigl[(\frac 1
{p!}\frac {\partial ^p f_{k-p}}{\partial \lambda
^p})(n)\bigr]_t\vspace{2mm}\\
\D = \sum_{p=1}^{k-1}\bigl(\frac 1 {p!}\frac {\partial ^p
f_{k-p}}{\partial \lambda ^p}\bigr)(n+\delta)=F_{k,s}(n+\delta).
 \ea
\] Therefore, the above function $F_{k,s} $ is a special
solution to the $k$-th subsystem.
 Then,
 it follows that
 the general solution to the $k$-th subsystem of
(\ref{eq:compactrepresentativesysytemforcomplexitonsol:toda-c}) is
given by
\[F_k=F_{k,h}+F_{k,s}= f_k+ \sum_{p=1}^{k-1}\frac 1 {p!}\frac {\partial
^p f_{k-p}}{\partial \lambda ^p}=  \sum_{p=0}^{k-1}\frac 1
{p!}\frac {\partial ^p f_{k-p}}{\partial \lambda ^p}.
\]
Finally, an induction ensures that the general solution to the
system
(\ref{eq:compactrepresentativesysytemforcomplexitonsol:toda-c}) is
given by
(\ref{eq:generalsolutionofcompactrepresentativesysytem:toda-c}).
 The proof is finished.
\end{proof}

Theorem \ref{thm:solutionformulaforeigenfunctions:toda-c} provides
us with an approach for solving a system of
differential-difference equations,
(\ref{eq:detrelationforn:toda-c}) and
(\ref{eq:detrelationfort:toda-c}), whose coefficient matrix
$\Lambda $ consists of Jordan blocks of the second type. Once we
solve the system of (\ref{eq:detrelationforn:toda-c}) and
(\ref{eq:detrelationfort:toda-c}), it follows from Theorem
\ref{thm:generalconditionsforCas:toda-c} that the corresponding
Casorati determinant gives us a solution to the Toda lattice
equation. If
 the coefficient matrix $\Lambda $ consists of
$m$ Jordan blocks of the second type in
(\ref{eq:2rdJordanblock:toda-c}), then the Casorati determinant
solution reads as
\begin{equation}\left\{\ba {l}
a_n=1+\D \frac{d^2}{dt^2}\log \tau_n,\
 b_n=\D \frac{d}{dt}\log \frac{ \tau _n}{\tau_{n+1}}
\,, \vspace{2mm}\\
\tau_n=\textrm{Cas}\bigl(\phi_1 (n),\cdots,\phi_{2l_1}(n);\cdots;
\phi _{2(l_1+\cdots +l_{m-1})+1} (n),\cdots,\phi_{2(l_1+\cdots
+l_m)}(n)\bigr)\,,\ea \right.
\label{eq:generalcomplexitonsolution:toda-c}
\end{equation}
where the involved eigenfunctions are determined by the formula
(\ref{eq:generalsolutionofcompactrepresentativesysytem:toda-c})
with $\lambda _j=\alpha_j+\beta _ji$, $1\le j\le m$.

In the following, we would like to show that the solutions defined
by (\ref{eq:generalcomplexitonsolution:toda-c}) are complexiton
solutions. To this end, let us write \be \lambda =2 \cosh \mu
=\textrm{e}^\mu +\textrm{e}^{-\mu },\ \mu \in \C
\label{eq:assumptionforlambda:toda-c} \ee and thus we can have \be
\omega = \textrm{e}^{ \mu },\ \mu \in \C .\ee Note that while $\mu
$ goes over the complex field, $\lambda =2 \cosh \mu  $ will
exhaust all complex values, and thus the assumption
(\ref{eq:assumptionforlambda:toda-c}) does not lose generality.
Then by Lemma \ref{le:gsof2ndtypesystem:toda-c}, the general
solution of the system
 (\ref{eq:gnlinearconditions:toda-c})
is given by \be f (n) = c \exp(  \mu  n+ t \exp( \delta \mu
))+d\exp( - \mu n+ t \exp(- \delta \mu ))
  , \label{eq:phi(mu)forgeneralnegaton:toda-c}
\ee  where $c$ and $d$ are arbitrary constants. The other
selection of $\omega =\textrm{e}^{-\mu }$ leads to the same
solution of the system
 (\ref{eq:gnlinearconditions:toda-c}).
Now write \be f=\phi_1+\phi_2i,\  \mu = a+bi,\ a\in \R ,\ b\ne
0\in \R ,\ee and assume that $c$ and $d$ are real constants but $
c^2+d^2\ne 0$ in order that $f\ne 0$. Then, we have \be
\lambda=\alpha +\beta i=2 (\cosh a\cos b)+2( \sinh a\sin b)i.\ee
Moreover, the system (\ref{eq:gnlinearconditions:toda-c}) becomes
the following system \be \left\{\ba {l} \phi_1(n+1)+\phi_1(n-1)= 2
(\cosh a\cos b) \phi_1(n) - 2( \sinh a\sin b)
\phi_2(n),\vspace{2mm}
\\
\phi_2(n+1)+\phi_2(n-1)= 2 (\sinh a\sin b) \phi_1(n) + 2 (\cosh
a\cos b )\phi_2(n),\vspace{2mm}
\\
(\phi_j(n))_t=\phi_j(n+\delta),\ j=1,2,\ea \right.
\label{eq:systemofphi_1andphi_2:toda-c}  \ee
  and its solution $(\phi_1,\phi_2)$ reads as \be
\left\{
\ba {l} \phi_1(n)=(\phi_1(a,b;c,d))(n):= \textrm{Re}(f(n))\vspace {2mm}\\
= c\textrm{e}^{ na+ t\textrm{e}^{\delta a}\cos \delta b}\cos( n b+
t\textrm{e}^{\delta a}\sin \delta b ) +d\textrm{e}^{- n a+
t\textrm{e}^{-\delta a}\cos \delta b }\cos( n b+
t\textrm{e}^{-\delta a}\sin \delta b )\,,
\vspace {2mm}\\
\phi_2(n)=(\phi_2(a,b;c,d))(n):= \textrm{Im}(f(n))\vspace {2mm}\\
= c\textrm{e}^{ n a+ t\textrm{e}^{\delta a}\cos \delta b}\sin( n
b+ t\textrm{e}^{\delta a}\sin \delta b ) -d\textrm{e}^{- n a+
t\textrm{e}^{-\delta a}\cos \delta b }\sin( n
b+t\textrm{e}^{-\delta a}\sin \delta b )\,.
 \ea \right. \label{eq:formulasforphi_1andphi_2:toda-c}\ee

The initial set (\ref{eq:detrelationforn:toda-c}) of difference
equations
 with the second type of Jordan
blocks [i.e., Jordan blocks in (\ref{eq:2rdJordanblock:toda-c})]
tells us that the solutions defined by
(\ref{eq:generalcomplexitonsolution:toda-c}) are associated with
the complex eigenvalues of the associated spectral problem. The
expressions (\ref{eq:formulasforphi_1andphi_2:toda-c}) and
(\ref{eq:generalsolutionofcompactrepresentativesysytem:toda-c})
for the required eigenfunctions
 indicate that the
resulting real solutions contain both exponential and
trigonometric functions of the space variable $n$. Therefore, it
follows that the solutions determined by
(\ref{eq:generalcomplexitonsolution:toda-c}) are real complexitons
to the Toda lattice equation (\ref{eq:toda:toda-c}), which
establishes the following theorem.

\begin{thm} Let $\alpha _j$ and $\beta _j$, $1\le j\le m$, be real
numbers and $\beta_j\ne 0,\ 1\le j\le m$. Then the Toda lattice
equation (\ref{eq:toda:toda-c}) has a class of real complexiton
solutions determined by the formula
(\ref{eq:generalcomplexitonsolution:toda-c}) and
 the formula
(\ref{eq:generalsolutionofcompactrepresentativesysytem:toda-c})
with $\lambda =\lambda _j=\alpha_j+\beta _ji$, $1\le j\le m$.
\end{thm}

A solution defined by (\ref{eq:generalcomplexitonsolution:toda-c})
is called an $m$-complexiton solutions (or simply, an
$m$-complexiton) of order $(l_1-1,l_2-1,\cdots,l_m-1) $ to the
Toda lattice equation (\ref{eq:toda:toda-c}). If $l_j=1,\ 1\le
j\le m$ or $m=1$, we simply say an $m$-complexiton solution or a
single complextion solution of order $l_1-1$. Based on the
expressions of eigenfunctions in
(\ref{eq:generalsolutionofcompactrepresentativesysytem:toda-c}),
 we see that the order $(l_1-1,l_2-1,\cdots,l_m-1) $ of
the complexiton reflect the maximum orders of derivatives of
eigenfunctions with respect to the corresponding eigenvalues.

\section{Construction of examples}

Theorems \ref{thm:generalconditionsforCas:toda-c} and
\ref{thm:solutionformulaforeigenfunctions:toda-c} provide a
general solution procedure using a general set of eigenfunctions.
It is, however, easier to apply some other techniques to present
concrete complexitons.

The simplest way to get a special set of eigenfunctions required
in complexitons is to take only one term in the expressions
(\ref{eq:generalsolutionofcompactrepresentativesysytem:toda-c})
for each $F_k$. This can be realized as follows. We set
\[ \Phi_2=(\phi_1,\phi_2)^T\] as before,
 and
assume that $f=\phi_1+\phi_2i$ solves the system
(\ref{eq:gnlinearconditions:toda-c}) with $\lambda =\alpha +\beta
i$. Then the choice of $c_k=d_k=0$, $2\le k\le l$, presents a
special set of eigenfunctions
 \be \bigl(\Phi_2^T(\lambda ),\frac 1{1!}
\partial _{\lambda } \Phi_2^T(\lambda ),\cdots, \frac 1{(l-1)!}
\partial _{\lambda }^{l-1} \Phi_2^T(\lambda )\bigr)^T
\label{eq:eigenfunctionsforsiglecomplexiton:toda-c} \ee to the
system of  (\ref{eq:detrelationforn:toda-c}) and
(\ref{eq:detrelationfort:toda-c}) with the coefficient matrix
$\Lambda $:
\[\Lambda = \left[\ba {cccc} A & & & 0
\vspace{2mm}\\
   I_2 & A& & \vspace{2mm}\\
\vdots & \ddots& \ddots& \vspace{2mm}\\
  0 &\cdots & I_2 &A \ea \right]_{l\times l},\ A=\left[\ba
{cc}\alpha &-\beta \vspace{2mm}\\ \beta &\alpha  \ea \right].
\]
Thus, a class of generalized Casorati determinant solutions to the
Toda lattice equation (\ref{eq:toda:toda-c}) can be generated from
the pair of eigenfunctions $\Phi_2=(\phi_1,\phi_2)^T$ as follows:
\begin{equation*}\left\{\ba {l}
a_n=1+\D \frac{d^2}{dt^2}\log \textrm{Cas}\bigl(\Phi_2 ^T(n),\D
\frac 1{1!}
\partial _{\lambda } \Phi _2^T(n),\cdots, \D \frac 1{(l-1)!}
\partial _{\lambda  } ^{l-1}\Phi _2^T(n)\bigr)\,,
\vspace{2mm} \\
 b_n=\D \frac{d}{dt}\log \frac{ \textrm{Cas}\bigl(\Phi _2^T(n),\D  \frac 1{1!}
\partial _{\lambda } \Phi_2 ^T(n),\cdots, \D \frac 1{(l-1)!}
\partial _{\lambda   } ^{l-1}\Phi _2^T(n)\bigr)}
{ \textrm{Cas}\bigl(\Phi _2^T(n+1), \D \frac 1{1!}
\partial _{\lambda  } \Phi _2^T(n+1),\cdots, \D \frac 1{(l-1)!}
\partial _{\lambda   } ^{l-1}\Phi _2^T(n+1)\bigr)
}\,. \ea \right.
\end{equation*}
A more general generalized Casorati determinant solution can be
constructed by combining pairs of eigenfunctions $\Phi_2(\lambda
_1),\Phi_2(\lambda _2),\cdots ,\Phi_2(\lambda _m)$ associated with
complex eigenvalues $\lambda _1,\lambda _2,\cdots ,\lambda _m$,
respectively. If the eigenvalues $\lambda _1,\lambda _2,\cdots
,\lambda _m$ have algebraic multiplicities $l_1,l_2,\cdots,l_m$,
respectively, the generalized Casorati determinant solution
generated from
\[\bigl(\Phi_2^T(\lambda _1),\cdots, \frac 1{(l_1-1)!} \partial
_{\lambda _1 }^{l_1-1} \Phi_2^T(\lambda _1);\cdots;
\Phi_2^T(\lambda _m),\cdots, \frac 1{(l_m-1)!} \partial _{\lambda
_m }^{l_m-1} \Phi_2^T(\lambda _m) \bigr )^T \] is an
$m$-complexiton solution of order $(l_1-1,l_2-1,\cdots,l_m-1) $ to
the Toda lattice equation (\ref{eq:toda:toda-c}). Here we clearly
see that the order $(l_1-1,l_2-1,\cdots,l_m-1) $ of the
complexiton is a sequence of the maximum orders of derivatives
with respect to the eigenvalues. The solution generated from the
set of eigenfunctions
(\ref{eq:eigenfunctionsforsiglecomplexiton:toda-c}) is a single
complexiton of order $l-1$.

However, it is not easy to compute the derivatives of
eigenfunctions with respect to eigenvalues. In what follows, to
avoid this difficulty, we would like to consider the system of
(\ref{eq:detrelationforn:toda-c}) and
(\ref{eq:detrelationfort:toda-c}) whose coefficient matrix
consists of the simplified blocks of the following type: \bea &&
\left[\ba {cccc} A_j & & & 0
\vspace{2mm}\\
   * & A_j& & \vspace{2mm}\\
\vdots & \ddots& \ddots& \vspace{2mm}\\
  * &\cdots &* &A_j \ea \right]_{l_j\times l_j},\ A_j=\left[\ba
{cc}\alpha _j&-\beta _j\vspace{2mm}\\ \beta _j&\alpha _j \ea
\right],
 \label{eq:2rdsimplifiedblock:toda-c} \eea
where $\alpha _j$ and $\beta _j>0$ are real constants, and the
symbol $*$ denotes an arbitrary entry. Since the Jordan forms of
matrices of this type are of the second type, the resulting
solutions are, of course, still complexitons. This form of the
coefficient matrix looks more complicated but it will bring us
convenience in computing concrete examples of complexitons.

Let us start from a set of eigenfunctions defined by
(\ref{eq:formulasforphi_1andphi_2:toda-c}).
 Taking
derivatives of $\Phi_2=(\phi_1,\phi_2)^T$ with respect to one of
the two constants $a$ and $b$ leads to
\begin{equation*}
S\left [\ba {c}\Phi_2 \vspace{2mm}\\
\frac 1{1!} \partial _{\xi } \Phi  _2\vspace{2mm}\\
\vdots \vspace{2mm}\\
\frac 1{(l-1)!} \partial _{\xi}^{l-1} \Phi_2
 \ea \right] =
\left [\ba {cccc} A  & & & 0\vspace{2mm}\\
\frac 1{1!} \partial
_{\xi } A  &A  & & \vspace{2mm}\\
\vdots & \ddots &\ddots & \vspace{2mm}\\
\frac 1{(l-1)!} \partial _{\xi}^{l-1} A & \cdots & \frac 1{1!}
\partial _{\xi} A &A \ea \right]_{l\times l} \left [
\ba {c}\Phi_2 \vspace{2mm}\\
\frac 1{1!} \partial _{\xi} \Phi _2 \vspace{2mm}\\
\vdots \vspace{2mm}\\
\frac 1{(l-1)!} \partial _{\xi}^{l-1} \Phi_2
 \ea \right]
,\end{equation*} and
\begin{equation*}
\bigl[(\frac 1{k!} \partial _{\xi  }^k \Phi_2 ) (n)\bigr]_{t}=
(\frac 1{k!}
\partial _{\xi }^k \Phi _2 )(n+\delta ), \ 0\le k\le l-1,
\end{equation*}
where $\xi$ denotes $a$ or $b$, $\partial _{\xi }$ is the
derivative with respect to $\xi$, and $A$ is defined by \be A=
\left[ \ba {cc} \alpha & -\beta \vspace{2mm} \\ \beta & \alpha \ea
\right]= \left[ \ba {cc} 2 \cosh a\cos b  & -2 \sinh a\sin b
\vspace{2mm}
\\ 2 \sinh
a\sin b  & 2  \cosh a\cos b  \ea \right]. \ee   This implies that
\be (\Phi_2^T,\frac 1{1!}
\partial _{\xi  } \Phi_2^T ,\cdots, \frac 1{(l-1)!} \partial _{\xi
}^{l-1} \Phi_2^T )^T\ee  is a special solution to the system \be
S\left[\ba{c} \phi_1\vspace{2mm}
\\ \phi_2 \vspace{2mm}
\\ \vdots \vspace{2mm}
\\ \phi_{2l-1}
\vspace{2mm}
\\ \phi_{2l}
 \ea \right]=
\left [\ba {cccc} A  & & & 0\vspace{2mm}\\
\frac 1{1!} \partial
_{\xi } A  &A  & & \vspace{2mm}\\
\vdots & \ddots &\ddots & \vspace{2mm}\\
\frac 1{(l-1)!} \partial _{\xi}^{l-1} A &\cdots & \frac 1{1!}
\partial _{\xi} A &A \ea \right]_{l\times l}
\left[\ba{c} \phi_1\vspace{2mm}
\\ \phi_2 \vspace{2mm}
\\ \vdots \vspace{2mm}
\\ \phi_{2l-1} \vspace{2mm}
\\ \phi_{2l}
 \ea \right], \label{eq:1stgeneraltriagularsystem:toda-c}\ee
 and \be
  \left[\ba{c} \phi_1(n)\vspace{2mm}
\\ \phi_2(n) \vspace{2mm}
\\ \vdots \vspace{2mm}
\\ \phi_{2l-1}(n) \vspace{2mm}
\\ \phi_{2l}(n)
 \ea \right]_t=\left[\ba{c} \phi_1(n+\delta )\vspace{2mm}
\\ \phi_2 (n+\delta )\vspace{2mm}
\\ \vdots \vspace{2mm}
\\ \phi_{2l-1}(n+\delta ) \vspace{2mm}
\\ \phi_{2l}(n+\delta )
 \ea \right],
\label{eq:2ndgeneraltriagularsystem:toda-c}\ee  the Jordan form of
whose coefficient matrix is of the second type.

 Therefore, for $\xi =a$
and $\xi =b$, we obtain two Casorati determinant
 solutions to the Toda lattice equation (\ref{eq:toda:toda-c}):
\begin{equation}\left\{\ba {l}
a_n=1+\D \frac{d^2}{dt^2}\log \textrm{Cas}\bigl(\Phi_2 ^T(n),\D
\frac 1{1!}
\partial _{\xi } \Phi _2^T(n),\cdots, \D \frac 1{(l-1)!}
\partial _{\xi  } ^{l-1}\Phi _2^T(n)\bigr)\,,
\vspace{2mm} \\
 b_n=\D \frac{d}{dt}\log \frac{ \textrm{Cas}\bigl(\Phi _2^T(n),\D  \frac 1{1!}
\partial _{\xi } \Phi_2 ^T(n),\cdots, \D \frac 1{(l-1)!}
\partial _{\xi  } ^{l-1}\Phi _2^T(n)\bigr)}
{ \textrm{Cas}\bigl(\Phi _2^T(n+1), \D \frac 1{1!}
\partial _{\xi } \Phi _2^T(n+1),\cdots, \D \frac 1{(l-1)!}
\partial _{\xi  } ^{l-1}\Phi _2^T(n+1)\bigr)
}\,, \ea \right.\label{eq:singlecomplexitonsol:toda-c}
\end{equation}
which correspond to the simplified blocks of the type in
(\ref{eq:2rdsimplifiedblock:toda-c}). Noting that $\phi_1$ and
$\phi_2$ are given explicitly by
(\ref{eq:formulasforphi_1andphi_2:toda-c}), it is direct to
compute the derivatives $\partial ^k_a\Phi_2$ and $\partial
^k_b\Phi_2$, $k\ge 0$, and further the corresponding complexitons.
Of course, from pairs of eigenfunctions
$(\phi_{1}(a_j,b_j),\phi_{2}(a_j,b_j))$, $ 1\le j\le m,$
 associated with different complex values $\mu _j=a_j+b_ji$,
$1\le j\le m$, two specific $m$-complexitons of order
$(l_1-1,l_2-1,\cdots,l_m-1) $ to the Toda lattice equation
(\ref{eq:toda:toda-c}) can be similarly constructed by taking
derivatives of the eigenfunctions with the involved pairs of two
constants $a_j$ and $b_j$.
 Two columns of
eigenfunctions required in those two complexitons are
\[ \bigl(\Phi_2^T(a_1,b_1), \cdots, \frac 1{(l_1-1)!}
\partial _{\xi _1 }^{l_1-1} \Phi_2^T(a _1,b_1);\cdots;
\Phi_2^T(a _m,b_m),\cdots, \frac 1{(l_m-1)!} \partial _{\xi _m
}^{l_m-1} \Phi_2^T(a _m,b_m) \bigr )^T , \] where $\xi_j=a_j$ or
$\xi_j =b_j$, $ 1\le j\le m,$ and
$\Phi_{2}(a_j,b_j)=(\phi_{1}(a_j,b_j),\phi_{2}(a_j,b_j))^T$, $
1\le j\le m,$ are defined by the formula
(\ref{eq:formulasforphi_1andphi_2:toda-c}) with $a=a_j$ and
$b=b_j$. This presents a large class of real complexitons to the
Toda lattice equation (\ref{eq:toda:toda-c}).

More generally, we can solve the the system of
(\ref{eq:1stgeneraltriagularsystem:toda-c}) and
(\ref{eq:2ndgeneraltriagularsystem:toda-c}) to get a broader set
of eigenfunctions required in complexitons. Let us still adopt \[
\lambda =\alpha +\beta i,\ F_k=\phi_{2k-1}+\phi_{2k}i,\  1\le k\le
l, \] as in Section \ref{sec:structureofcomplexitons:toda-c},
 then the system of (\ref{eq:1stgeneraltriagularsystem:toda-c}) and
(\ref{eq:2ndgeneraltriagularsystem:toda-c}) is equivalent to the
following triangular system for all $F_k$: \be SF_{k} =\lambda
F_{k}+\sum_{p=1}^{k-1}\frac 1 {p!}\frac {\partial ^p \lambda
}{\partial \xi ^p}F_{k-p}, \ (F_{k}(n))_t=F_{k} (n+\delta ) ,\
1\le k\le l, \label{eq:compactgeneraltriagularsystem:toda-c}
 \ee where $F_0=0$. This system contains the system
 (\ref{eq:compactrepresentativesysytemforcomplexitonsol:toda-c})
 as a special case.

\begin{thm}\label{thm:solutionformulaforgeneraltriagularsystem:toda-c}
 Let $\lambda =\lambda (\xi)$ be a function from $\C $ to $\C -\{2\}$ and $\delta =\pm 1$.
 Suppose that $f(\lambda ;c,d)$ is the general solution to (\ref{eq:gnlinearconditions:toda-c}), defined by
(\ref{eq:generalsolforgnlinearconditions:toda-c}), and for $1\le
k\le l$, define $f_k=f(\lambda (\xi ) ;c_k,d_k)$ with a pair of
arbitrary constants $c_k$ and $d_k$. Then the general solution to
the general triangular system of differential-difference equations
(\ref{eq:compactgeneraltriagularsystem:toda-c}) is given by \be
F_k=\sum_{p=0}^{k-1}\frac 1 {p!} \frac {\partial ^p
f_{k-p}}{\partial \xi ^{p}} =\sum_{p=0}^{k-1}\frac 1 {p!} \frac
{\partial ^p f(\lambda(\xi);c_{k-p},d_{k-p})}{\partial \xi ^{p}} ,
\ 1\le k\le l.
\label{eq:generalsolutionofcompactgeneraltriagularsystem:toda-c}
\ee
\end{thm}

\begin{proof} The proof is similar to the one of Theorem
\ref{thm:solutionformulaforeigenfunctions:toda-c}.
 Note that the general solution expression (\ref{eq:generalsolutionofcompactgeneraltriagularsystem:toda-c})
 can be rewritten as
\[F_k=F_{k,h}+F_{k,s},\ F_{k,h}=f_k,\ F_{k,s}=
\D \sum_{p=1}^{k-1}\frac 1 {p!}\frac {\partial ^p
f_{k-p}}{\partial \xi ^p},\ 1\le k\le l.
\]
By Lemma \ref{le:gsof2ndtypesystem:toda-c}, $F_{k,h}=f_k$ is the
general solution to the homogeneous counterpart of the $k$-th
subsystem for $F_k$ in
(\ref{eq:compactgeneraltriagularsystem:toda-c}). Use the same
argument as in the proof of Theorem
\ref{thm:solutionformulaforeigenfunctions:toda-c}, what we need to
prove now is that $F_{k,s}$ is a special solution to the $k$-th
subsystem for $F_k$ in
(\ref{eq:compactgeneraltriagularsystem:toda-c}).

 At the current situation, from
\[ Sf_k=\lambda (\xi)   f_k, \ (f_k(n))_t=f_k(n+\delta ),\ 1\le k\le l,\]
we have
  \be \left\{\ba {l}
\D S(\frac 1 {p!} \frac {\partial ^p f_{k}}{\partial \xi ^{p}}) =
\sum_{q=0}^p \frac 1 {q!}  \frac {\partial ^{q} \lambda (\xi)
}{\partial \xi ^{q}}\frac 1 {(p-q)!} \frac {\partial ^{p-q}
f_{k}}{\partial
\xi ^{p-q}},\  1\le k\le l, \ p\ge 1,\vspace{2mm}\\
 \D  \bigl[(\frac 1 {p!} \frac {\partial ^p f_{k}}{\partial
\xi ^{p}})(n)\bigr]_t=(\frac 1 {p!} \frac {\partial ^p
f_{k}}{\partial \xi ^{p}})(n+\delta ) ,\ 1\le k\le l,\  p\ge 1,
\ea \right.
\label{eq:ktimesderivativeofF_kingeneraltriagularsystem:toda-c}
\ee the former equality of which is equivalent to
 \be
(S-\lambda (\xi)   )(\frac 1 {p!} \frac {\partial ^p
f_{k}}{\partial \xi ^{p}}) = \sum_{q=1}^p \frac 1 {q!}  \frac
{\partial ^{q} \lambda (\xi)   }{\partial \xi ^{q}} \frac 1
{(p-q)!} \frac {\partial ^{p-q} f_{k}}{\partial \xi ^{p-q}},\ 1\le
k\le l,\ p\ge
1.\label{eq:equiktimesderivativeofF_kingeneraltriagularsystem:toda-c}\ee
Therefore, using
(\ref{eq:equiktimesderivativeofF_kingeneraltriagularsystem:toda-c}),
we can compute that
\[ \ba {l}
\D (S-\lambda (\xi)   )F_{k,s}=\sum_{p=1}^{k-1} (S-\lambda (\xi)
) \frac 1 {p!}\frac {\partial ^p f_{k-p}}{\partial \xi ^p}
\vspace{2mm}\\
\D  = \sum_{p=1}^{k-1} \sum_{q=1}^p \frac 1 {q!}\frac {\partial
^{q}\lambda (\xi)   }{\partial \xi ^{q}} \frac 1 {(p-q)!}\frac
{\partial ^{p-q}f_{k-p}}{\partial \xi ^{p-q}}
\vspace{2mm}\\
\D  = \sum_{q=1}^{k-1} \sum_{p=q}^{k-1} \frac 1 {q!}\frac
{\partial ^{q}\lambda (\xi)   }{\partial \xi ^{q}} \frac 1
{(p-q)!}\frac {\partial ^{p-q}f_{k-p}}{\partial \xi ^{p-q}}
\vspace{2mm}\\
\D  = \sum_{q=1}^{k-1} \frac 1 {q!}\frac {\partial ^{q}\lambda
(\xi)  }{\partial \xi ^{q}}  \sum_{p=q}^{k-1} \frac 1
{(p-q)!}\frac {\partial ^{p-q}f_{k-p}}{\partial \xi ^{p-q}}
\vspace{2mm}\\
\D  = \sum_{q=1}^{k-1} \frac 1 {q!}\frac {\partial ^{q}\lambda
(\xi)  }{\partial \xi ^{q}} \sum_{p=0}^{(k-q)-1} \frac 1 {p!}\frac
{\partial ^pf_{(k-q)-p}}{\partial \xi ^{p}}
\vspace{2mm}\\
\D =\sum_{q=1}^{k-1}\frac 1 {q!}\frac {\partial ^{q}\lambda (\xi)
}{\partial \xi ^{q}} F_{k-q},\ea
\] and by using the latter equality of
(\ref{eq:ktimesderivativeofF_kingeneraltriagularsystem:toda-c}),
we have
\[ \ba {l}
(F_{k,s}(n))_t= \D \sum_{p=1}^{k-1}  \bigl[(\frac 1 {p!}\frac
{\partial ^p f_{k-p}}{\partial \xi ^p})(n)\bigr]_t
\vspace{2mm}\\
\D =\sum_{p=1}^{k-1}  (\frac 1 {p!}\frac {\partial ^p
f_{k-p}}{\partial \xi ^p})(n+\delta ) = F_{k,s}(n+\delta ). \ea \]
 Therefore, the above function $F_{k,s} $ is a special
solution to the $k$-th subsystem of
(\ref{eq:compactgeneraltriagularsystem:toda-c}), indeed.
 The proof is finished.
\end{proof}

This theorem provides a general set of eigenfunctions required in
complexitons, while using
(\ref{eq:1stgeneraltriagularsystem:toda-c}) and
(\ref{eq:2ndgeneraltriagularsystem:toda-c}). Now it is just a
direct computation to construct a complexiton solution from a
$\tau$-function $\tau_n$.

In particular, we can start from the pair of eigenfunctions
$\phi_1$ and $\phi_2$ defined by
(\ref{eq:formulasforphi_1andphi_2:toda-c}) to compute examples of
complexitons. First without computing derivatives of $\phi_1$ and
$\phi_2$, the $\tau$-function of a single complexiton can be
expressed as
\begin{eqnarray}
 \tau_n &=&\textrm{Cas}(\phi_1(n),\phi_2(n))\nonumber \\
 & =&
2cd \,\textrm{e}^{2t\cosh \delta a \cos \delta b
}\sin(2nb+ b+2t\cosh \delta a \sin \delta b  )\sinh a \nonumber \\
&& + c^2\textrm{e}^{2na+a+2t\textrm{e}^{\delta a}\cos \delta b
}\sin b-d^2\textrm{e}^{-2na-a+2t\textrm{e}^{-\delta a}\cos
 \delta b }\sin b\,,\label{eq:taufunctionforsinglecomplexiton:toda-c}\end{eqnarray}
 where $\delta=\pm 1$ and $a,b,c,d$ are arbitrary real constants, but
$b\ne 0$ and $c^2+d^2\ne 0$ in order that $\tau_n\ne 0$. If we fix
$c=\pm d$, and the $\tau$-function boils down to
\begin{eqnarray}
 \tau_n &=&2c^2 \, \textrm{e}^{2t\cosh \delta a \cos \delta b }\sinh
(2na+a+2t\sinh \delta a \cos \delta b)\sin b \nonumber  \\ && \pm
2c^2 \,\textrm{e}^{2t\cosh \delta a \cos \delta b }\sin(2nb+
b+2t\cosh \delta a \sin \delta b  )\sinh a \,.
\label{eq:simplifiedtaufunctionforsinglecomplexiton:toda-c}\end{eqnarray}
Second, through computing the first-order derivatives of $\phi_1$
and $\phi_2$, the $\tau$-function of a single complexiton of order
1 reads as \be \tau_n=\textrm{Cas}\bigl(\phi_1(n),\phi_2(n),\frac
{\partial \phi_1(n)}{\partial \xi},\frac {\partial
\phi_2(n)}{\partial
\xi}\bigr),\label{eq:simplifiedtaufunctionforcomplexitonof1storder:toda-c}
\ee where $\xi=a$ or $\xi =b$. More generally, upon choosing
arbitrary real constants $a_i,b_i$, $i=1,2$ and $c_i,d_i$, $1\le
i\le 3$, which satisfy $b_i\ne 0$, $i=1,2$ and $c_i^2+d_i^2\ne 0$,
$1\le i\le 3$, we can have a $\tau$-function of a single
complexiton of order 1:
\begin{eqnarray} &&
\tau_n=\textrm{Cas}\bigl((\phi_1(a_1,b_1;c_1,d_1))(n),(\phi_2(a_1,b_1;c_1,d_1))(n),
\nonumber \\
&& \qquad (\phi_1(a_1,b_1;c_2,d_2))(n)+\frac {\partial
(\phi_1(a_1,b_1;c_1,d_1))(n)}{\partial \xi_1},\nonumber
\\ &&\qquad (\phi_2(a_1,b_1;c_2,d_2))(n)+\frac {\partial
(\phi_2(a_1,b_1;c_1,d_1))(n)}{\partial \xi_1}\bigr),
\label{eq:1stmoregeneralsimplifiedtaufunctionforcomplexitonof1storder:toda-c}
\end{eqnarray}
and a $\tau$-function of a 2-complexiton of order (1,1):
\begin{eqnarray} &&
\tau_n=\textrm{Cas}\bigl((\phi_1(a_1,b_1;c_1,d_1))(n),(\phi_2(a_1,b_1;c_1,d_1))(n),
\nonumber \\
&& \qquad \frac {\partial (\phi_1(a_1,b_1;c_1,d_1))(n)}{\partial
\xi _1},
\frac {\partial (\phi_2(a_1,b_1;c_1,d_1))(n)}{\partial \xi_1},\nonumber \\
 &&
 \qquad (\phi_1(a_2,b_2;c_2,d_2))(n),(\phi_2(a_2,b_2;c_2,d_2))(n),
\nonumber\\
&&\qquad  (\phi_1(a_2,b_2;c_3,d_3)(n)+ \frac {\partial
(\phi_1(a_2,b_2;c_2,d_2))(n)}{\partial \xi_2} ,\nonumber \\
&& \qquad  (\phi_2(a_2,b_2;c_3,d_3)(n)+\frac {\partial
(\phi_2(a_2,b_2;c_2,d_2))(n)}{\partial \xi_2}
\bigr),\label{eq:2ndmoregeneralsimplifiedtaufunctionforcomplexitonof1storder:toda-c}
\end{eqnarray}
where $\xi_i=a_i$ or $\xi_i =b_i$, $i=1,2$.

\section{Concluding remarks}

A set of coupled conditions consisting of differential-difference
equations has been proposed for Casorati determinants to solve the
Toda lattice equation. A systematic analysis has been made for
solving the resulting system of differential-difference equations
whose coefficient matrix consists of Jordan blocks of the second
type, together with the solution formula for the key subsystem
associated with one Jordan block. The resulting set of
eigenfunctions leads to complexitons through the Casoratian
formulation. Moreover, a feasible way has been presented to
construct sets of eigenfunctions required for forming
complexitons, which allows us to directly compute examples of real
complexitons.

We remark that the resulting complexitons of order zero (i.e., not
involving derivatives of eigenfunctions) can be constructed from
complexification of wave numbers of $2$-solitons (see
\cite{Jaworski-PLA1984} for the KdV case). However, the resulting
complexitons of order larger than zero (i.e., involving
derivatives of eigenfunctions) can not be generated from
complexification of solitons. Such solutions are generated on the
basis of our coupled conditions established in Theorem
\ref{thm:generalconditionsforCas:toda-c}. On the other hand, based
on Theorem \ref{thm:generalconditionsforCas:toda-c}, our
generalized Casorati determinant solutions give solitons and
negatons if $\lambda >2$, positons if $\lambda <2$ and rational
solutuons to the Toda lattice equation if $\lambda =2$
\cite{MarunoMO-JPSJ2003,MaY-CSF2004}. Viewing $(S-2)\phi$ as a
discrete version of $\partial _x^2\phi$, we can easily see that
this is consistent with the phenomenon in the KdV case
\cite{MaY-TAMS2003}.

Our results also indicate that integrable equations can have three
different kinds of explicit exact transcendental function
solutions: negatons, positons and complexitons. Solitons are
usually a specific class of negatons. Roughly speaking, negatons
and positons are solutions which involve exponential functions and
trigonometric functions of space variables, respectively, and they
are all associated with real eigenvalues of the associated
spectral problems. But complexitons are different solutions which
involve both exponential and trigonometric functions of space
variables, and they are associated with complex eigenvalues of the
associated spectral problems. Interaction solutions among
negatons, positons, rational solutions and complexitons are a
class of much more general and complicated solutions to soliton
equations, in the category of elementary function solutions. There
is also a large class of $\theta$-function solutions to soliton
equations. It is an interesting question for us what inverse
scattering data there exist for complexitons of the Toda lattice
equation.

It is also natural to ask whether our idea of constructing
complexitons can be successfully applied to other integrable
lattice equations such as the Ablowitz-Ladik (AL) equation
\cite{AblowitzL-JMP1976,AblowitzC-book1991} and general Toda
lattice equations \cite{KodamaY-PD1996}. Particularly interesting
to us is to make an extension to full discrete integrable
equations such as the discrete-time KdV equation
\cite{Hirota-JPSJ1977KdV} and the discrete-time Toda lattice
equation \cite{Hirota-JPSJ1977Toda}. On the other hand, it has
been pointed out that multi-positon solutions of the KdV equation
may be related to giant ocean waves such as ``freak wave'' (rogue
wave), breaking up ships \cite{Matveev-TMP2002}. It is our hope
that complexitons can provide certain mathematical background for
related nonlinear phenomena in the field of mathematical physics.

\vskip 2mm \noindent {\bf Acknowledgments:}
 The authors acknowledge supports from the University of South Florida Internal Awards Program
(Grant No. 1249-936RO), the Faculty Development Grant Program of
the College of Arts and Sciences of the University of South
Florida, and the 21st Century COE program ``Development of Dynamic
Mathematics with High Functionality'' at Faculty of Mathematics,
Kyushu University.

\end{document}